\documentclass{article}

\usepackage[utf8]{inputenc}
\usepackage[T1]{fontenc}
\usepackage{lmodern}
\usepackage{textcomp}
\usepackage{amsmath, amssymb}
\usepackage{graphicx}
\usepackage{tikz}
\usepackage{mathtools}
\usepackage{venndiagram}
\usepackage{braket}
\usepackage{arabtex}
\usepackage{utf8}

\setcode{utf8}
\newcommand{\noon}{\RL{ن}}

\title{Potential for Polynomial Solution for NP-Complete Problems using\\Quantum Computation}
\author{Neema Rustin Badihian}
\date{26th of April, 2025}

\begin{document}

\maketitle

\section{Abstract}

In this paper, we propose two new methods for solving Set Constraint Problems, as well as a potential polynomial solution for NP-Complete problems using quantum computation. While current methods of solving Set Constraint Problems focus on classical techniques, we offer both a quantum-inspired matrix method and a quantum matrix method that neutralizes common contradictions and inconsistencies that appear in these types of problems. We then use our new method to show how a potential polynomial solution for NP-Complete problems could be found using quantum computation. We state this as a potential solution, rather than an actual solution, as the outcome of any quantum computation may not be the same as the expected outcome. We start by formally defining a Set Constraint Problem. We then explain current, classical methods that are used to solve these problems and the drawbacks of such methods. After this, we explain a new quantum-inspired matrix method that allows us to solve these problems, with classical limitations. Then, we explain a new quantum matrix method that solves these problems using quantum information science. Finally, we describe how we can extend this method to potentially solve NP-Complete problems in polynomial time using quantum computation.

\section{Definition of Set Constraint Problem}
To define a Set Constraint Problem (SCP), we start with a Universal set that may contain all sets, a series of subsets that are present in the Universal set, and a series of constraints that describe the relationship between a given element and a given subset of the Universal set:

\begin{align*}
U = & \{x_1, x_2, \cdots, x_n\}\\
S = & \{S_1, S_2, \cdots, S_m\}\\
C = & \{c_1, c_2, \cdots, c_k\}
\end{align*}

With $U$, $S$, and $C$ defined, we can now define the problems themselves. We define a Set Constraint Problem as a collection of logical constraints involving elements and subsets of a Universal set $U$. The SCP generalizes the classical Set Cover Problem, which was proven to be NP-complete in Karp's foundational work~\cite{karp1972}. There are three main types of constraints that we are interested in. They are as follows:

\begin{align*}
\text{Element Inclusion: } & x \in S_i \\
\text{Element Exclusion: } & x \notin S_i \\
\text{Element Difference: } & x \in S_i \setminus S_j \\
\end{align*}

The objective of an SCP is to find assignments for each set $S_i \subseteq U$ such that all given constraints are satisfied.

\section{Classical Methods}
There are four main classical methods to solve a Set Constraint Problem (SCP). These are: the Venn Diagram method, the Symbolic Deduction method, the Boolean Logic method, and the Binary Table method. We now explain each method formally.

\subsection{Venn Diagram Method}
In the Venn Diagram method, we create a circle for each set $S_i$ that we are trying to define. The space that these circles rest within represents the Universal set $U$. We arrange the circles so that all possible intersections between the sets are visually represented.\\

Constraints are then applied, one element at a time. When a constraint indicates that an element $x$ belongs to a particular set $S_i$, the element is placed within the circle that corresponds to the set $S_i$. If $x$ is found to belong to multiple sets ($S_1, S_2, \cdots, S_n$), it is placed in the region that corresponds to the intersection of the sets ($S_1 \cap S_2 \cap \cdots \cap S_n$).\\

This is repeated for each given constraint. If any elements are known to be in the Universal set $U$, but are not described with enough information in the constraints, they are either not added to the diagram or they are placed in the space outside the circles, indicating they are in the Universal set but not in any of the subsets $S_i$ of $U$. \\

As a result, this method requires much trial and error and often fails to explain the full relationship between the sets and the elements. It can work well for a small number of sets (3 or fewer), but once we start having a larger number of sets, it can become very difficult to use. Furthermore, it does not account for any uncertainty in regards to the elements. For example, an element may be constrained such that it could (rather than should) belong to two sets, but not to a third, yet the diagram provides no way to reflect this partial information. These limitations highlight the need for a more systematic approach that can scale to larger problems, handle ambiguity, and enforce constraints through logical inference rather than visual approximation.

\subsection{Symbolic Deduction Method}
The Symbolic Deduction method attempts to solve SCPs by translating all constraints into formal logical propositions, after which, we use inference to deduce the inclusion or exclusion of elements in the sets. Just as with the Venn Diagram method, we start with a Universal set $U$, a series of sets $(S_1, S_2, \cdots, S_n)$ within $U$, and a list of constraints. Each constraint is reinterpreted as precise logical statements. Below is a series of constraints and the results of the reinterpretations:

\begin{align*}
x \in S_i & \Rightarrow \text{Record that } x \in S_i \\
x \notin S_i & \Rightarrow \text{Record that } x \notin S_i \\
x \in S_i \setminus S_j & \Rightarrow \text{Record that } x \in S_i \text{ and } x \notin S_j \\
\end{align*}

Now that we have rephrased the constraints into logical rules, we are able to deduce new facts through inference. We do this by: identifying new facts that follow the existing known ones, applying logic rules to derive conclusions, and tracking dependencies and identity conflicts. We repeat this process until we have gone through all constraints. \\

This method, similarly to the Venn Diagram method, does not account for uncertainty. With this method, we determine an element either is in a set, or not in a set. We leave no room for ambiguity. This method also involves very little written work. It mostly relies on the memory of the problem solver. It does not follow a specific system that allows for keeping track of the information we receive from the constraints; instead, it relies on inference. As such, contradictions that may appear are only discovered reactively. We may deduce that one element is in a specific set, only to find later that it may not be in that set. Lastly, unlike the Venn Diagram method, it makes use of no visual tool that could help us solve the problem. In short, it may be viewed as an incomplete method for solving SCPs.

\subsection{Boolean Logic Method}
The Boolean Logic method attempts to solve SCPs by assigning a boolean variable to every pair of elements and sets. For all elements $\{x_1, x_2, \cdots, x_n\}$ of the Universal set $U$, each one is paired with a set from our set of sets $S = \{S_1, S_2, \cdots, S_k\}$, as such: $(x_i, S_j)$. We then assign a boolean variable that represents the relationship between the element and the set:

\[
B_{i,j} = 
\begin{cases}
1 & \text{if } x_i \in S_j \\
0 & \text{if } x_i \notin S_j
\end{cases}
\]

We can express our three main types of constraints using this Boolean Logic method:
\begin{align*}
\text{Inclusion: } & x_i \in S_j \Rightarrow B_{i, j} = 1 \\
\text{Exclusion: } & x_i \notin S_j \Rightarrow B_{i, j} = 0 \\
\text{Element Difference: } & x \in S_i \setminus S_j \Rightarrow (B_{x, i} = 1) \land (B_{x, j} = 0) \\
\end{align*}

With our new boolean definitions for these types of constraints, we then apply these constraints to the existing variables in the problem we are given. This uses the same logical inference that was required in Symbolic Deduction, with no visual representation such as that of the Venn diagram method.\\

Despite appearing more formal than our previous methods, this method is still deterministic and leaves no room for ambiguity. It also requires backtracking, as we might find that previous assumptions are false after we formalize a boolean variable for a new constraint in any given problem. It requires that we keep track of much of the information in our minds, without using a formalized method to keep track as we progress through the problem.

\subsection{Binary Table Method}

The Binary Table method is closely related to the Boolean Logic method. It assigns values of $0$ or $1$ for each possible element in each possible set. It differs primarily in that it uses a table or matrix to keep track of these values directly.\\

As with previous methods, it leaves no room for uncertainty. If an element's relationship to a set remains undefined, with neither $0$ nor $1$ being placed in the corresponding cell, we are left with no formal method on how to proceed. Instead, we either claim we do not know enough about the element, make a guess as to which set the element could be in, or we default to $0$ to say that the element is not in the set. None of these options are grounded in the method itself.

\section{Quantum-Inspired Matrix Method (QIMM)}

All previously discussed methods for solving SCPs require a binary choice for determining whether an element is in a set, given a series of constraints. In the classical methods, either an element is in a set, or it is not in a set. This leaves no room for ambiguity or uncertainty.\\

Our Quantum-Inspired Matrix method introduces a third logical value to represent uncertainty. Rather than limiting set membership to the binary choices of $0$ and $1$, we instead define three possible values. The third value adds a level of uncertainty, representing cases where we cannot say definitively whether an element is in a set.\\

To express this state more explicitly in symbolic notation, we introduce the Persian letter $\noon$ as a new operator representing unresolved membership. We choose this symbol both for its visual clarity and for its symbolic heritage as the curve of the letter may be viewed as a wave and the dot may be viewed as a fish jumping out of the wave. We find that this suits our needs well when considering the behavior particles, which exist in probabilistic waves prior to measurement. Putting the possible symbolic interpretation of the letter together with our current understanding of quantum mechanics, we may view the fish as a particle or possibility emerging from the wave of probability. When using this symbol, it may be read as "could be in", in contrast with $\in$ which may be read as "in" and $\notin$ which may be read as "not in". Using this new notation, we have the following values for our matrix:

\[
\begin{cases}
1 & \text{if } x_i \in S_j \\
0 & \text{if } x_i \, \noon \, S_j \\
-1 & \text{if } x_i \notin S_j
\end{cases}
\]

We represent these different states in a matrix $M \in \{-1, 0, 1\}^{n\times k}$, where each row corresponds to an element $x_i \in U$ and each column corresponds to a final set $S_j$. The entry $M(x_i, S_j)$ captures the membership relationship between an element and a set as defined by our ternary logic. We express the three main types of constraints that we are interested in with our new system:

\begin{align*}
\text{Element Inclusion: } & x_i \in S_j \Rightarrow M(x_i, S_j) \coloneqq 1 \\
\text{Element Exclusion: } & x_i \notin S_j \Rightarrow M(x_i, S_j) \coloneqq -1 \\
\text{Element Difference: } & x_i \in S_p \setminus S_q \Rightarrow M(x_i, S_p) \coloneqq 1 \land M(x_i, S_q) \coloneqq -1\\
\end{align*}

At this point, we demonstrate this methodology with an example. Consider an SCP with the following properties:

\begin{align*}
U = & \{a, b, c, d, e, f, g\} \\
S = & \{X, Y, Z\}\\
C = & \begin{cases}
    X \setminus Y = \{a, d\}\\
    X \setminus Z = \{d\}\\
    Y \setminus X = \{b, f\}\\
    Z \setminus X = \{b\}\\
    Z \setminus Y = \{a\}\\
    c \notin X\\
    e \in Z
    \end{cases}
\end{align*}

where $U$ is the Universal set containing the elements $a$ through $g$, $S$ is a series of subsets of $U$, and $C$ is a series of constraints that describe the subsets listed in $S$. If we apply the given constraints, the following matrix is revealed:

\begin{table}[h]
  \centering
  \begin{tabular}{|c|c|c|c|}
    \hline
    & X & Y & Z \\ \hline
    a & 1 & -1 & 1 \\ \hline
    b & -1 & 1 & 1 \\ \hline
    c & -1 & 0 & 0 \\ \hline
    d & 1 & -1 & -1 \\ \hline
    e & 0 & 0 & 1 \\ \hline
    f & -1 & 1 & 0 \\ \hline
    g & 0 & 0 & 0 \\ \hline
  \end{tabular}
  \label{tab:matrix}
\end{table}

The first five constraints each describe element differences. For example, in the first constraint, we are told that when the elements of $Y$ are removed from $X$, the remaining elements in $X$ are $a$ and $d$. The final two constraints describe a set exclusion constraint and a set inclusion constraint. After going through each of the given constraints, we find that several of the relationships between elements and sets lack a description. We place zeros in each matrix cell where we do not have enough information, in order to state that the element could be in ($\noon$) the set, but we are uncertain.\\

With our matrix filled out completely, we can now describe each of the subsets of $U$. For the entries with a value of $1$, we know to include the element in the respective set. For the entries with a value of $-1$, we know to exclude the element from the respective set. As for the entries with a value of $0$, the choice of including or excluding the element from the respective set is slightly ambiguous, similarly to the data that the zeros represent. In the context of classical computation, more data leads to more accurate results. If the problem solver desired to maximize their accuracy, they may create multiple versions of each set to describe all possibilities. In our example, the solver could create four separate sets to describe the set $X$. These could include:\\

\[
    \text{$X$-0, which could represent $X$ excluding both $e$ and $g$.}
\]
\[
    \text{$X$-1, which could represent $X$ excluding $e$ and including $g$.}
\]
\[
    \text{$X$-2, which could represent $X$ including $e$ and excluding $g$.}
\]
\[
    \text{$X$-3, which could represent $X$ including both $e$ and $g$.}
\]

Including these versions of $X$ would increase our accuracy; however, it would also increase the amount of data. With larger problems and with more uncertainty, this could potentially result in an unwieldy amount of data to compute. For each set that has $u$ uncertain elements, we would require $2^{|u|}$ versions to solve it. As such, we leave the choice of which version(s) to include to the problem solver, although we offer some considerations. We caution that when choosing to include any version(s) of a specific set, the problem solver may want to consider the balance between the version(s) of the set they have chosen, as well as the version(s) of the other sets they may choose to include. Choosing to include both a version of $X$ and a version of $Y$ with $e$, may result in an overrepresentation of element $e$ in the data. It may also result in a favoring of either $X$ or $Y$, depending on how including $e$ affects the results of the computation. In short, choosing versions without considering balance may result in inaccurate data being presented. To get around this, we offer two options: we may choose to leave out the uncertain data altogether, or we may choose to pursue a quantum computation.

\section{Quantum Matrix Method (QMM)}

The Quantum Matrix method involves taking our classical data, representing it as quantum data, and solving the given SCP using quantum computation. To represent our classical data as quantum data, we define the following:

\[
\begin{cases}
\ket{0} & \text{if } x_i \in S_j \\
\frac{1}{\sqrt{2}}(\ket{0} + \ket{1}) & \text{if } x_i \, \noon \, S_j \\
\ket{1} & \text{if } x_i \notin S_j
\end{cases}
\]

We may represent the Universal set as a series of additions, rather than a series of separate individual elements. Similarly, the subsets of the Universal set may also be represented as a series of additions.

\begin{align*}
U = & \{\ket{0} \cdot (a + b + c + d + e + f + g)\}\\
S = & \{\ket{0} \cdot (X + Y + Z)\}
\end{align*}

When we use these definitions with our example problem, the following matrix is revealed:

\begin{table}[h]
  \centering
  \begin{tabular}{|c|c|c|c|}
    \hline
    & X & Y & Z \\ \hline
    a &$ \ket{0}$ & $ \ket{1}$ & $ \ket{0}$\\ \hline
    b &$ \ket{1}$ & $ \ket{0}$ & $ \ket{0}$\\ \hline
    c &$ \ket{1}$& $\frac{1}{\sqrt{2}}(\ket{0} + \ket{1})$ & $\frac{1}{\sqrt{2}}(\ket{0} + \ket{1})$ \\ \hline
    d &$ \ket{0}$ & $ \ket{1}$ & $ \ket{1}$\\ \hline
    e & $\frac{1}{\sqrt{2}}(\ket{0} + \ket{1})$ & $\frac{1}{\sqrt{2}}(\ket{0} + \ket{1})$ &$ \ket{0}$\\ \hline
    f &$ \ket{1}$ & $ \ket{0}$& $\frac{1}{\sqrt{2}}(\ket{0} + \ket{1})$ \\ \hline
    g & $\frac{1}{\sqrt{2}}(\ket{0} + \ket{1})$ & $\frac{1}{\sqrt{2}}(\ket{0} + \ket{1})$ & $\frac{1}{\sqrt{2}}(\ket{0} + \ket{1})$ \\ \hline
  \end{tabular}
  \label{tab:matrix}
\end{table}

\newpage

We may now define each set formally, using the information in this matrix:

\[
X = \ket{0} \cdot (a + d) + \ket{1} \cdot (b + c + f) + \frac{1}{\sqrt{2}}(\ket{0} + \ket{1}) \cdot (e + g)
\]
\[
Y = \ket{0} \cdot (b + f) + \ket{1} \cdot (a + d) + \frac{1}{\sqrt{2}}(\ket{0} + \ket{1}) \cdot (c + e + g)
\]
\[
Z = \ket{0} \cdot (a + b + e) + \ket{1} \cdot d + \frac{1}{\sqrt{2}}(\ket{0} + \ket{1}) \cdot (c + f + g)
\]

With this, we have formally solved the Set Constraint Problem using a quantum method within the scope of classical solutions. We purposely leave the uncertain entries in an ambiguous state, as the information given in the problem lacks certainty for these entries.

\section{Quantum Matrix Method with Separate States (QMMSS)}

The elements in the Universal set may each be represented by a separate state created by a separate quantum particle. For a Universal set with $n$ elements, we would require $n$ particles to represent each state. We can then prepare each particle as either $\ket{0}$ when the corresponding element is in the set, $\ket{1}$ when the element is not in the set, or $\frac{1}{\sqrt{2}}(\ket{0} + \ket{1})$ when the element could be in the set. If we are given $m$ sets with $n$ possible elements in each set, we would need a total of $m \times n$ particles to represent each possible state. To prepare each set would take constant time of $O(1)$ for each state, therefore, it would take a total polynomial time of $O(m \times n)$. For these reasons, measuring each particle for state collapse would also take polynomial time of $O(m \times n)$. The only issue we see here is that the actual outcome of wave collapse is not necessarily the same as the expected outcome, and as such, it may require multiple rounds of computation to output the expected result.\\

The Set Constraint Problem is a known NP-Complete problem. With this method, we are able to potentially solve any membership problem in polynomial time using quantum computing, if not for the fact that the actual outcome may differ from the expected outcome. Furthermore, as all NP-Complete problems may be viewed as membership problems~\cite{cook1971}, we have shown that any NP-Complete problem may potentially be solved in polynomial time using quantum computing with the Quantum Matrix Method with Separate States.

\section{Conclusion}

The final solution of the QIMM leaves room for interpretation for the problem solver. For QMM, the final solution leaves purposeful ambiguity, as is the nature of quantum descriptions. Finally, the QMMSS method shows that all NP-Complete problems can potentially be solved in polynomial time using quantum computation. Uncertainty can be challenging to represent using classical methods, but quantum methods provide the necessary tools. The Quantum-Inspired Matrix Method and the Quantum Matrix Method provide more thorough solutions for SCPs than those of existing classical methods. The Quantum Matrix Method with Separate States brings a new potential contribution to addressing the uncertainty surrounding solutions to NP-Complete problems.


\newpage

\begin{thebibliography}{9}

\bibitem{karp1972}
R.~M.~Karp, 
\textit{Reducibility Among Combinatorial Problems}, 
in R.~E.~Miller and J.~W.~Thatcher (eds.), \textit{Complexity of Computer Computations}, Springer, 1972, pp.~85--103.\\

\bibitem{cook1971}
S.~A.~Cook,
\textit{The Complexity of Theorem-Proving Procedures}
in \textit{Proceedings of the Third Annual ACM Symposium on Theory of Computing (STOC)}, 1971, pp.~151--158.

\end{thebibliography}
\end{document}